\documentclass[aps, prb, citeautoscript, twocolumn]{revtex4-2} 

\usepackage{amsmath} 
\usepackage{amsfonts}
\usepackage{graphicx}
\usepackage{braket}
\usepackage{cancel}
\setcitestyle{super,open={},close={}}

\begin{document}

\title{Complementarity and entanglement in a simple model of inelastic scattering}\thanks{Forthcoming from the \textit{American Journal of Physics}.}

\author{David Kordahl}
\email{dkordahl@centenary.edu} 
\affiliation{Department of Physics and Engineering, Centenary College of Louisiana, Shreveport, LA 71104}
\date{\today}

\begin{abstract}
A simple model coupling a one-dimensional beam particle to a one-dimensional harmonic oscillator is used to explore complementarity and entanglement. This model, well-known in the inelastic scattering literature, is presented under three different conceptual approaches, with both analytical and numerical techniques discussed for each. In a purely classical approach, the final amplitude of the oscillator can be found directly from the initial conditions. In a partially quantum approach, with a classical beam and a quantum oscillator, the final magnitude of the quantum-mechanical amplitude for the oscillator's first excited state is directly proportional to the oscillator's classical amplitude of vibration. Nearly the same first-order transition probabilities emerge in the partially and fully quantum approaches, but conceptual differences emerge. The two-particle scattering wavefunction clarifies these differences and allows the consequences of quantum entanglement to be explored.
\end{abstract}

\maketitle 

\section{Introduction}

An understanding of quantum concepts is often built by overlapping classical analogies, analytical models, and numerical illustrations. After learning about a particle’s classical momentum, students are shown that a particle's de Broglie wavelength depends on the inverse of that momentum. Students may then analytically model the one-dimensional (ID) reflection and transmission of de Broglie waves from a potential barrier, whose analogous classical counterparts would all have been stopped by that same barrier. Further insights can be gained by numerically modeling such reflection and transmission events using wave packets. Each of these approaches teases out new qualitative and quantitative connections \cite{modern_physics}.

Recent years have seen an increasing consensus that the concept of entanglement—the inability of some quantum states to be written as the product of individual particle states—should be a part of every student’s quantum tookit. As Daniel V. Schroeder pointed out in ``Entanglement isn’t just for spins,” entanglement generically arises when quantum particles interact with each other \cite{entanglement_not_spins}. In that paper, Schroeder presented two dynamical models showing how entanglement emerges, but lamented that such examples are rarely included in quantum mechanics textbooks. ``The reason,” he conceded, ``is probably that despite their conceptual simplicity, a quantitative treatment of either scenario requires numerical methods.”

This paper presents a conceptually simple model that can model entanglement without resorting to numerical methods. The level of mathematical difficulty in this treatment is similar to that of the commonly taught models involving potential barriers. The model is a simplified 1D treatment of inelastic scattering. It is well-known to the electron spectroscopy community \cite{original_reference}, and is similar to the model of a  1D atom scattering off a 1D harmonic oscillator presented in this journal several decades ago by Knudson \cite{Knudson1975}, though this treatment differs in its attention to the time-evolution of the scattering process.

\begin{figure}
\begin{center}
  \includegraphics[width=.6\columnwidth]{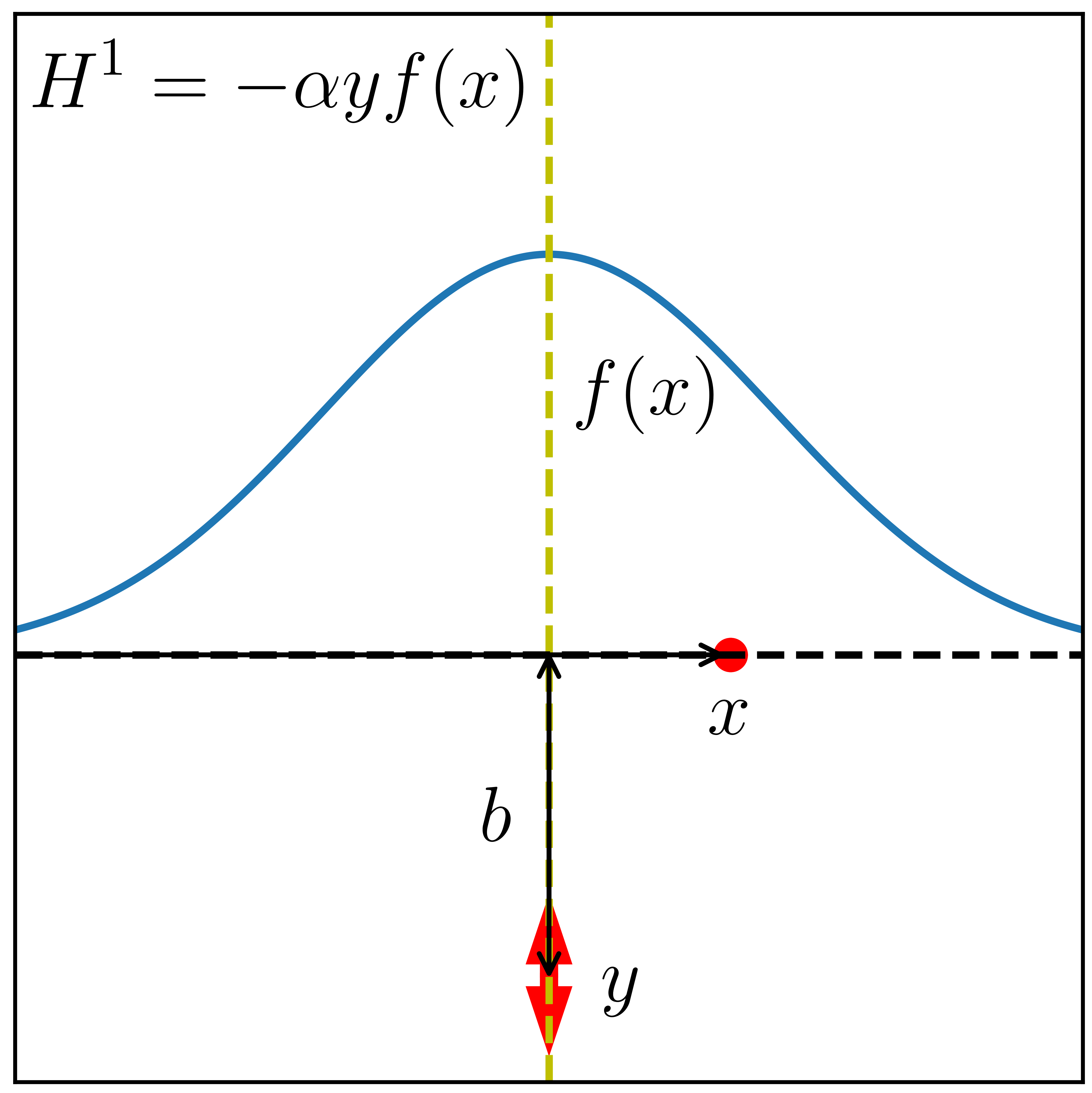}
\end{center}
  \caption{In the model being studied, one variable ($x$) represents the position of a beam particle, and the other ($y$) represents the vibrational displacement of an oscillator. The beam and oscillator are coupled via $H^1$, a potential that depends on the product of a coupling constant $\alpha$, the oscillator displacement $y$, and some spatially-dependent function $f(x)$, which may depend implicitly on an impact parameter $b$.}\label{fig:model illustration}
\end{figure}

In the model under review, a beam particle is coupled to a harmonic oscilator, as illustrated in Fig.~\ref{fig:model illustration}. The model Hamiltonian sums the contributions of a 1D free particle of mass $m$ (position variable $x$, conjugate momentum $p_x$), a 1D harmonic oscillator of reduced mass $\mu$ and resonance frequency $\omega_0$ (position variable $y$, conjugate momentum $p_y$), and an interaction term $H^1$ that couples the two systems:
\begin{align}\label{eq:model Hamiltonian}
H &=  &H^0_\mathrm{beam} &+& H^0_\mathrm{HO} \qquad &+& H^1 \quad \nonumber \\
&= &\underbrace{\frac{p_x^2}{2m}}_\text{beam} &+ &\underbrace{\frac{p_y^2}{2 \mu} + \frac{1}{2} \mu \omega_0^2 y^2}_\text{harmonic oscillator} \,&-&\underbrace{\alpha \, y \, f(x)}_\text{{interaction}}.
\end{align}
The coupling constant $\alpha$ rationalizes units and tunes the strength of the interaction, and the spatially-varying ``window function" $f(x)$ is taken to have units of length$^{-2}$, and to die off as $x$ approaches $\pm \infty$. The negative sign in $H^1$ means that the oscillator is attracted toward the beam when $\alpha$ and $f(x)$ are positive, but changing this sign would not significantly alter any results.

Though its form is simple, this model can be used to capture real physics. For instance, in ``Characterizing Localized Surface Plasmons Using Electron Energy-Loss Spectroscopy,'' Cherqui et al.\cite{EELS plasmons} derive ``classical" and ``quantum" Hamiltonians for the electron-plasmon interaction (their Eqs.~14 and 15) that map respectively onto our Eq.~\ref{eq:model Hamiltonian} and Eq.~\ref{eq:oscillator hamiltonian - raising/lowering}, but for the fact that an electron couples to infinitely many plasmonic modes, while our beam couples to just one oscillator. For the plasmonic case, as an electron nears a nanoparticle, a surface charge is induced, which in turn interacts with the electron via the Coulomb potential. The physics in this model is analogous, with the passing beam tugging on the oscillator, and the oscillator in turn tugging back on the beam. 

In addition to its discussion of entanglement, this presentation also highlights conceptual issues surrounding classical/quantum complementarity.\cite{complementarity} Complementarity is explored implicitly as different approaches to the model are presented---a classical approach (Sec.~\ref{sec:classical}), and, after a review of perturbation theory (Sec.~\ref{sec:perturbation theory}), partially and fully quantum approaches (Secs.~\ref{sec:partially quantum approach} and~\ref{sec:fully quantum approach}). Complementarity then is addressed explicitly in Sec.~\ref{sec:complementarity}, where the transition from one approach to the next is discussed, and Sec.~\ref{sec:entanglement} explores post-interaction entanglement, showing how predictions may change in entangled systems following a measurement. Sec.~\ref{sec:conclusion} is a short summary.

\section{Classical Approach}\label{sec:classical}

Solving the model in the classical approach requires only standard tools. Applying Hamilton's equations
\begin{align}
\frac{dx}{dt} &= \frac{\partial H}{\partial p_x} & \,\frac{dy}{dt} &= \frac{\partial H}{\partial p_y} \\
\frac{dp_x}{dt} &= -\frac{\partial H}{\partial x} & \, \frac{dp_y}{dt} &= -\frac{\partial H}{\partial y}
\end{align}
to Eq.~\ref{eq:model Hamiltonian} yields the classical equations of motion for the beam position $x$ and oscillator displacement $y$
\begin{equation}\label{eq:classical equations of motion}
\begin{split}
m \frac{d^2 x}{dt^2} &= \alpha y \frac{df}{dx} \\
\mu \frac{d^2 y}{dt^2} &= -\mu \omega_0^2 y + \alpha f(x)
\end{split}
\end{equation}
which can be solved either by analytical or numerical methods.

\subsection{Analytical Calculation}

If we suppose that the kinetic energy of our beam far exceeds the magnitude of the interaction energy between the beam and the harmonic oscillator, we will be safe in approximating the motion of the beam particle as 
\begin{equation}
x \approx vt.
\end{equation}
Under this approximation, the equation of motion for the oscillator becomes that of a driven harmonic oscillator:
\begin{equation}\label{eq:classical driven oscillator}
\mu \frac{d^2 y}{dt^2} = -\mu \omega_0^2 y + \alpha f(vt).
\end{equation}

Using the Fourier transform convention
\begin{equation}\label{eq:temporal FT}
\tilde{f}(\omega) = \frac{1}{\sqrt{2\pi}} \int_{-\infty}^{+\infty} dt e^{i \omega t} f(vt),
\end{equation}
we can transform both sides of Eq.~\ref{eq:classical driven oscillator} to obtain
\begin{equation}
-\mu \omega^2 \tilde{y}(\omega) = -\mu \omega_0^2 \tilde{y}(\omega) + \alpha \tilde{f}(\omega),
\end{equation}
which can easily be solved algebraically to give us
\begin{equation}\label{eq:classical FT of y}
\tilde{y}(\omega) = \frac{-\alpha \tilde{f}(\omega)}{\mu (\omega^2 - \omega_0^2)}.
\end{equation}

In principle, the story that this tells is simple. As the beam passes, it pulls on the oscillator a bit, and once the beam is far enough away, the oscillator will be left vibrating with whatever amplitude it had once the beam and oscillator were sufficiently separated to effectively decouple. And in principle, we should be able to find the final amplitude of the oscillator's vibration by performing the inverse Fourier transform of Eq.~\ref{eq:classical FT of y} to find $y(t)$.

In practice, however, we would like our calculation to depend less sensitively on phases, so instead we can calculate the \textit{work} done on the beam. This work will be negative, since the beam loses energy as it passes, but the magnitude of this energy loss equals the magnitude of the energy gain by the oscillator, since the combined system is energy-conserving. This transfer can then be used to calculate the oscillator amplitude.

To calculate the work done on the beam, one can (a) rewrite the work integral as an integral over time; (b) insert the expression for $y(t)$ as an inverse Fourier transform; (c) reverse the order of the time and frequency integrals; and (d) perform the frequency integral by slightly displacing the poles off the real axis by $i \epsilon$ and using the residue theorem. These steps are carried out in detail in the Appendix. This gives us the result that 
\begin{equation}\label{eq:work on beam}
W_\mathrm{beam} =-\frac{\pi \alpha^2}{\mu} \left| \tilde{f}(\omega_0) \right|^2
\end{equation}

The work done on the oscillator by the beam will have the same magnitude, but with the opposite sign:
\begin{equation}\label{eq:work on oscillator}
W_\mathrm{HO} = \frac{\pi \alpha^2}{\mu} \left| \tilde{f}(\omega_0) \right|^2.
\end{equation}
We might notice that this work done on the oscillator is proportional to to the oscillator amplitude squared. If the oscillator's classical amplitude is $y_m$, we can write
\begin{equation}
W_\mathrm{HO} = \frac{1}{2} \mu \omega_0^2 y_m^2, 
\end{equation}
which in turn gives us that 
\begin{equation}\label{eq:general classical amplitude}
y_m = \frac{\sqrt{2\pi} \alpha}{\mu \omega_0} \left|\tilde{f}(\omega_0)\right|.
\end{equation}

For specificity, let's consider an example. Suppose we approximate a dipole potential with a thinner-tailed Gaussian \cite{dipole_pot}, such that our window function $f(x)$ is
\begin{equation}\label{eq:window function}
f(x) = b^{-2} e^{-x^2/b^2}.
\end{equation}
If we insert our approximate solution $x \approx vt$ into this
\begin{equation}
f(vt) = b^{-2} e^{-v^2 t^2/b^2},
\end{equation}
we can take the Fourier transform (Eq.~\ref{eq:temporal FT}) to yield
\begin{equation}
\tilde{f}(\omega) = \frac{e^{-b^2\omega^2/4 v^2}}{\sqrt{2}b v},
\end{equation}
so the work done on the oscillator (Eq.~\ref{eq:work on oscillator}) is 
\begin{equation}
W_\mathrm{HO} = \frac{\pi \alpha^2}{2\mu v^2 b^2 } e^{-b^2\omega_0^2/2 v^2}.
\end{equation}
and the oscillator's vibrational amplitude (Eq.~\ref{eq:general classical amplitude}) is
\begin{equation}\label{eq:specific classical amplitude}
y_m = \frac{\sqrt{\pi} \alpha}{\mu \omega_0} \frac{e^{-b^2 \omega_0^2/4 v^2}}{v b}.
\end{equation}

\begin{figure}
\begin{center}
  \includegraphics[width=\columnwidth]{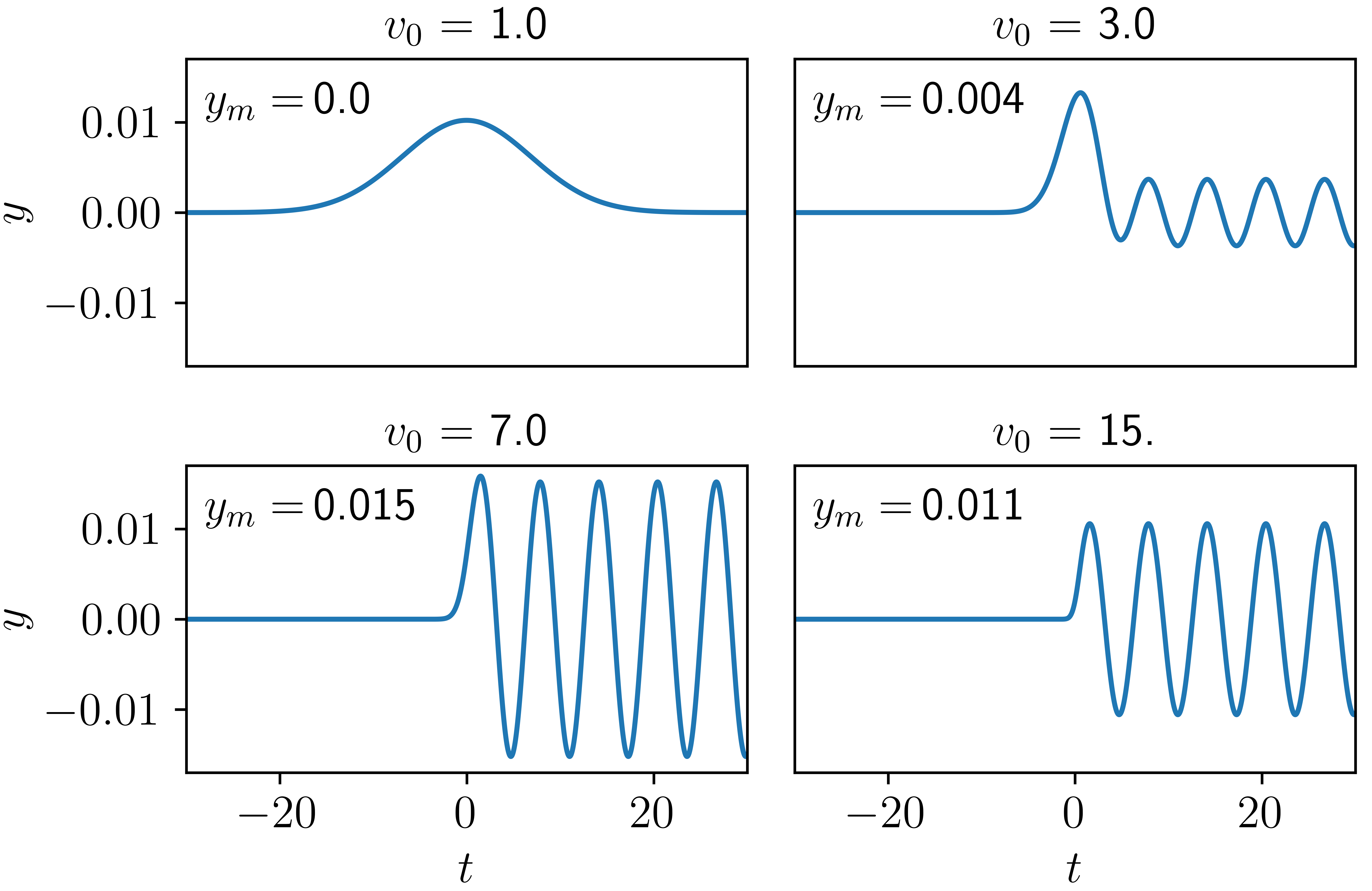}
\end{center}
  \caption{Time-evolution of a classical oscillator ($y$) for different initial speeds of passing beam ($x$): 
  $v_0 = 1.0$ (top left), $v_0 = 3.0$ (top right), $v_0 = 7.0$ (bottom left), and $v_0 = 15$ (bottom right). The final amplitude $y_m$ is given for each case.}\label{fig:classical outcome}
\end{figure}

\subsection{Numerical Calculation}

Of course, one may avoid Fourier transforms altogether and simply evolve the equations of motion numerically, using, for instance, the Euler-Richardson method \cite{comp_phys}. Using units where $\hbar = \omega_0 = m = 1$, $b = 10$, and $\mu = 100$, and using the window function specified in Eq.~\ref{eq:window function}, the results of this are shown in Fig.~\ref{fig:classical outcome} for four different initial velocities of the beam particle (i.e., of $dx/dt$ at $t \ll -b/v$), of $v$ equal to 1.0, 3.0, 7.0, and 15. In each of the subplots, the classical amplitude $y_m$ has been checked numerically, and matches the value predicted analytically by Eq.~\ref{eq:specific classical amplitude}.

\section{Perturbation Theory Review}\label{sec:perturbation theory}

But there is a problem. The classical approach is empirically inadequate for micro- or mesoscopic systems, since the beam particle in fact will not lose energy each time it passes the (generalized) oscillator. Sometimes the beam particle will be observed to lose energy, but most of the time it will pass by without any energy loss at all.

As a result, we will want to develop a quantum version of the model. We are interested in both analytical and numerical solutions, and when we want analytical solutions in quantum mechanics, we often turn to perturbation theory. So let's briefly remind ourselves, now, about time-dependent perturbation theory \cite{perturbation_theory}. 

Suppose we have an unperturbed Hamiltonian operator $\hat{H_0}$ whose eigenvectors $\ket{n^0}$ are known:
\begin{equation}\label{eq:unpurturbed eigenstates}
\hat{H^0} \ket{n^0} = E_n^0 \ket{n^0}.
\end{equation} 
Solutions of the Schrodinger equation 
\begin{equation}\label{eq:Schrodinger equation}
i \hbar \frac{\partial \ket{\psi}}{\partial t} = \hat{H} \ket{\psi}
\end{equation}
can be written, without loss of generality, in terms of coefficients $c_n(t)$:
\begin{equation}\label{eq:expansion of psi in eigenstates}
\ket{\psi(t)} = \sum_n c_n(t) e^{-i E_n^0 t/\hbar} \ket{n^0}.
\end{equation}

When the Hamiltonian in question is the unperturbed Hamiltonian $\hat{H} = \hat{H^0}$, one can substitute Eq.~\ref{eq:expansion of psi in eigenstates} into the Schrodinger equation to confirm that the coefficients $c_n$ are constant in time. When the Hamiltonian is perturbed by a weak (and possibly time-dependent) term $\hat{H^1}(t)$ such that $\hat{H} = \hat{H^0} + \hat{H^1}(t)$, the $c_n$ coefficients will still be useful, since they vary more slowly in time than state coefficients would without factoring out $\exp(-i E_n^0 t/\hbar)$. 

Applying the operator $\bra{f^0} \exp(+i E_f^0 t/\hbar)$ to both sides of Eq.~\ref{eq:Schrodinger equation} and inserting the expansion of Eq.~\ref{eq:expansion of psi in eigenstates}, we can find the time-dependence of any state coefficient $c_f$ as
\begin{equation}\label{eq:time-dependent expansion}
i \hbar \frac{d c_f}{dt} = \sum_n \bra{f^0} H^1(t) \ket{n^0} e^{i \omega_{fn} t}
\end{equation}
where
\begin{equation}
\omega_{fn} = \frac{E_f^0 - E_n^0}{\hbar}.
\end{equation}

For a system that begins in initial state $\ket{i^0}$ with energy $E_i$ at time $t = -\infty$, Eq.~\ref{eq:time-dependent expansion} leads to a first-order perturbation coefficient $c_f$ for exciting the system to some final state $\ket{f^0}$ with energy $E_f$ as
\begin{equation}\label{eq:first-order perturbation}
c_{f}(t) = \delta_{fi} - \frac{i}{\hbar} \int_{-\infty}^{t} \bra{f^0} H^1(t') \ket{i^0} e^{i \omega_{fi} t'} dt',
\end{equation}
which means the first-order probability of transition to state $\ket{f^0}$ is found by allowing $t \rightarrow \infty$ and calculating 
\begin{equation}\label{eq:Born rule}
P_f = |c_f(t = +\infty)|^2
\end{equation}
from the Born rule. When the initial and final states are distinct, inserting Eq.~\ref{eq:first-order perturbation} into Eq.~\ref{eq:Born rule} yields
\begin{equation}\label{eq:first-order probabilities, general}
P_f = \frac{1}{\hbar^2} \left|\int_{-\infty}^{+\infty} \bra{f} \hat{H^1}(t') e^{i \omega_{fi} t'} \ket{i} dt' \right|^2,
\end{equation}
which will be sufficient to let us calculate excitation probabilities analytically for both of the common quantum-mechanical approaches to the model being reviewed. 

\section{Partially Quantum Approach}\label{sec:partially quantum approach}

The partially quantized approach has us treat the oscillator as quantized while treating the beam only as the source of a time-dependent perturbation. If the beam is still modeled as a classical particle whose path follows $x \approx vt$, the Hamiltonian that acts on the oscillator wavefunction is
\begin{equation}\label{eq:oscillator hamiltonian partial}
\hat{H} = \underbrace{\frac{p_y^2}{2 \mu} + \frac{1}{2}\omega_0^2 \mu y^2}_{\centering \text{$\hat{H^0}$}} - \underbrace{\alpha y f(v t)}_{\centering \text{$\hat{H^1}(t)$}}.
\end{equation}
This form will allow us to use perturbation theory, since the energy eigenstates of the quantum harmonic oscillator $\ket{n}$, with $E_n = (n+1/2)\hbar \omega_0$, are well-known.

\subsection{Analytical Calculation}

Of course, for the quantum harmonic oscillator $y$ and $p_y$ do not commute, but follow
\begin{equation}\label{eq:commutator}
[y, p_y] = y p_y - p_y y = i \hbar.
\end{equation}
The typical move, now, is to rewrite the Hamiltonian in terms of creation and annihilation operators
\begin{equation}
\begin{split}
a^\dagger &= \frac{1}{\sqrt{2\hbar\omega_0 \mu}} \left(\omega_o \mu y - i p_y \right) \\
a &= \frac{1}{\sqrt{2\hbar\omega_0 \mu}} \left(\omega_o \mu y + i p_y \right),
\end{split}
\end{equation}
which recasts the reduced Hamiltonian (Eq.~\ref{eq:oscillator hamiltonian partial}) as
\begin{equation}\label{eq:oscillator hamiltonian - raising/lowering}
H = \underbrace{\hbar \omega_0 \left(a^\dagger a + \frac{1}{2} \right)}_\text{$\hat{H_0}$} - \underbrace{\alpha f(v t)\sqrt{\frac{\hbar}{2 \omega_0 \mu}}  \left( a + a^\dagger \right)}_\text{$\hat{H^1}(t)$}.
\end{equation}

If the quantized oscillator begins its ground state of $\ket{0}$ with energy $E_0 = \hbar \omega_0/2$, we can calculate its probability of being kicked into its excited energy eigenstate $\ket{n}$ at energy $E_n = (n+1/2) \hbar \omega_0$ using Eqn.~\ref{eq:first-order probabilities, general}:
\begin{equation}
P_n = \frac{1}{\hbar^2} \left|\int_{-\infty}^{+\infty} \bra{n} \hat{H^1}(t) e^{i n \omega_0 t} \ket{0} dt \right|^2
\end{equation}
This expression predicts that the only possible transition (considering first-order perturbations) is from $\ket{0} \rightarrow \ket{1}$, with the probability
\begin{equation}\label{eq:P1, partially quantum}
P_1 = \frac{\pi \alpha^2}{\hbar \mu \omega_0} |\tilde{f}(\omega_0)|^2.
\end{equation}

We might pause, now, to reflect on how this compares to the outcome of the purely classical system. In the purely classical system, we found that the final amplitude of the oscillator $y_m$ could be calculated deterministically as a function of the the initial beam speed. For the partially quantum system, the same can be said of the probability $P_1$ of finding the quantum harmonic oscillator in its first excited state. 

In fact, if we compare $y_m$ and $P_1$, we find that 
\begin{equation}
P_1 = \frac{\mu \omega_0}{2 \hbar} y_m^2.
\end{equation}
or, equivalently, that
\begin{equation}\label{eq:connection between classical amplitude and quantum probability}
y_m = \sqrt{\frac{2 \hbar}{\mu \omega} P_1}
\end{equation}
To belabor this point, we notice that the probability $P_1$ is proportional to the square of the magnitude of the coefficient to $\ket{1}$, so, to first order, the classical oscillation amplitude is directly proportional to the magnitude of the quantum amplitude of the $\ket{0} \rightarrow \ket{1}$ transition.

\subsection{Numerical Calculation}\label{subsec:partially quantum numerics}

\begin{figure}
\begin{center}
  \includegraphics[width=\columnwidth]{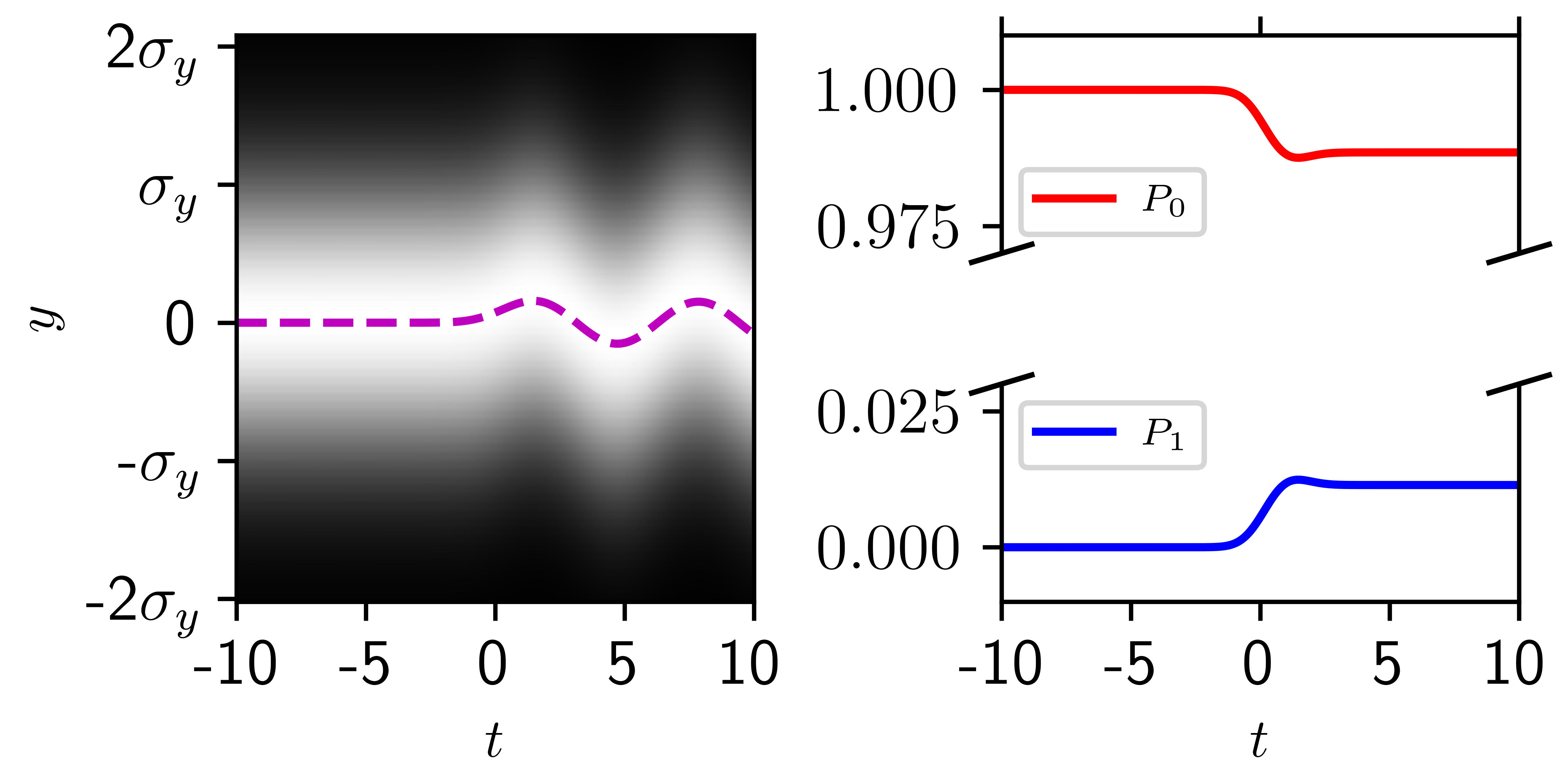}
\end{center}
  \caption{A classical beam ($v_0 = 7.0$) is coupled to a quantum oscillator. \textit{Left}: $\psi^*(y)\psi(y)$ for the quantum harmonic oscillator in the partially quantum approach is shown as a function of $t$ in black and white, and the expectation value for the oscillator's displacement, $\langle y \rangle$, is overlaid as a dashed line. \textit{Right}: The probabilities of measuring the oscillator in state $\ket{0}$ (top) or $\ket{1}$ (bottom) are shown as a function of $t$.}\label{fig:partially quantum}
\end{figure}

What about a straightforward numerical solution? Suppose we begin with the oscillator in its ground state, represented as the normalized position wavefunction
\begin{equation}\label{eq:oscillator normalized ground state}
\psi_0(y) = \braket{y|0} = \left(\frac{1}{\pi \sigma_y^2} \right)^{1/4} e^{- y^2/2 \sigma_y^2},
\end{equation}
and we want to know its probability of transitioning to its first excited state
\begin{equation}\label{eq:oscillator normalized first excited state}
\psi_1(y) = \braket{y|1} = \frac{\sqrt{2}}{\sigma_y} \left(\frac{1}{\pi \sigma_y^2} \right)^{1/4} y \, e^{-y^2 /2 \sigma_y^2}
\end{equation}
where in both $\psi_0(y)$ and $\psi_1(y)$
\begin{equation}
\sigma_y = \sqrt{\frac{\hbar}{\mu \omega_0}}.
\end{equation}

The most direct way to proceed is simply to use the time-dependent Schrodinger equation
\begin{equation}
i \hbar \frac{\partial}{\partial t} \psi(y) = \hat{H}(t) \psi(y)
\end{equation}
where in the position representation the time-dependent Hamiltonian operator looks like
\begin{equation}
\hat{H}(t) = -\frac{\hbar^2}{2 \mu} \frac{\partial^2}{\partial y^2} + \frac{1}{2}\mu \omega_o^2 y^2 - \alpha y f(v t).
\end{equation}
If the system starts in its ground state at some time $t \ll -b/v$, then it be can evolved forward in time using finite-difference time-domain (FDTD) methods \cite{integrate_PDE}.  

The numerical calculation for this is shown in Fig.~\ref{fig:partially quantum}, for $v = 7$ and the other model parameters kept the same as for the classical simulations, as quoted in Sec.~\ref{subsec:partially quantum numerics}. The dashed purple trace in the figure is the numerically calculated expected value of the oscillator displacement
\begin{equation}\label{eq:y expected value}
\langle y \rangle = \int_{-\infty}^{+\infty} dy \, \psi^{*}(y) \, y \,\psi(y).
\end{equation}
which, as we might expect, follows the classical result $y(t)$ for $v=7$ shown in Fig.~\ref{fig:classical outcome}, as will be further established in our discussion of complementarity below (Sec.~\ref{sec:complementarity}).

From there, one can also calculate the overlap between the wavefunction and its various energy eigenstates to find probabilities. One may numerically calculate
\begin{equation}
\braket{\psi_1|\psi} = \int_{-\infty}^{+\infty} dy \, \psi_1^*(y) \psi(y)
\end{equation}
to find the probability of transition from $\ket{0} \rightarrow \ket{1}$ as
\begin{equation}
P_1 = |\braket{\psi_1|\psi}|^2,
\end{equation}
and similarly to find the probability of remaining in state $\ket{0}$ as $P_0$. These probabilities are shown on the right in Fig.~\ref{fig:partially quantum}, and they match the analytical prediction of Eq.~\ref{eq:P1, partially quantum}.

\section{Fully Quantum Approach}\label{sec:fully quantum approach}

If we want to treat our model in a fully quantum-mechanical way, the simplest way is to include the beam and the oscillator in a combined quantum state. 

To make the problem analytically tractable, it will be useful to introduce a box length $L$ over which our wavefunction runs in $x$, so as to make it normalizable. This will allow us to smuggle in classical notions like the velocity of the particle, since the time integral as the beam goes from $-L/2$ to $+L/2$ can be written, if we consider that the beam travels at a speed of approximately $v$ throughout, as going from $t_{-}=-L/2v$ to $t_{+} = +L/2v$. 

Plane waves constitute energy eigenfunctions of the free particle, which we can write in terms of $k$-vectors with $k = mv/\hbar$. We can also reuse the energy eigenstates of the harmonic oscillator, which we will now express as $\ket{n}_y$. So we can write a general quantum-mechanical state for the beam-oscillator system  as
\begin{equation}
\begin{split}
&\ket{\psi(x,y,t)}  \\
&\,\,= \sum_{k_x,n} c_{k_x,n}(t) \left(\frac{e^{i k_x x}}{\sqrt{L}} e^{-i E_k t/\hbar} \right) \left(\ket{n}_y e^{-i E_n t/\hbar}\right)
\end{split}
\end{equation}
where the $\sqrt{L}$ in the denominator is for plane-wave normalization, and the energies are just
\begin{equation}
\begin{split}
E_k &= \frac{\hbar^2 k_x^2}{2 m}, \\
E_n &= \left(n + 1/2\right) \hbar \omega_0.
\end{split}
\end{equation}

\subsection{Analytical Calculation}

The Hamiltonian operator in the position basis is
\begin{equation}\label{eq:hamiltonian position basis full quantum}
\begin{split}
\hat{H} = &-\frac{\hbar^2}{2 m} \frac{\partial^2}{\partial x^2} -\frac{\hbar^2}{2 \mu} \frac{\partial^2}{\partial y^2} + \frac{1}{2}\mu \omega_o^2 y^2 \\
&- \alpha y f(x).
\end{split}
\end{equation}
Products of the plane waves along $x$ and harmonic oscillator states along $y$ are energy eigenstates of this combined Hamiltonian, but for the coupling term linking the two:
\begin{equation}
H^1 = - \alpha y f(x).
\end{equation}
Using this, we can proceed here just as above, calculating the transition coefficient $c_{k_x,n}$ as $t\rightarrow\infty$ using Eq.~\ref{eq:first-order perturbation}. 

The initial state $\ket{i^0}$, with energy $E_i$, can be written as
\begin{equation}\label{eq:initial state fully quantum}
\begin{split}
\ket{i^0} &= \frac{e^{i k_0 x}}{\sqrt{L}} \ket{0}_y \\
E_i &=\frac{\hbar^2 k_0^2}{2 m} + \hbar \omega_0/2,
\end{split}
\end{equation}
and the final state $\ket{f^0}$, with energy $E_f$, as
\begin{equation}
\begin{split}
\ket{f^0} &= \frac{e^{i k_1 x}}{\sqrt{L}} \ket{n}_y \\
E_f &=\frac{\hbar^2 k_1^2}{2 m} + \left(n+1/2 \right) \hbar \omega_0.
\end{split}
\end{equation}
It is worth noticing that while the energy eigenstates of the oscillator are fairly well-localized in $y$, the plane-wave energy eigenstates of the beam are spread out over the entire space in $x$. Though these plane-wave beam states connect only loosely to ``particle'' concepts, their status as energy eigenstates of the uncoupled beam allows us to use standard perturbation theory.

At this point, we can once again use Eq.~\ref{eq:first-order probabilities, general} to calculate our first-order transition probabilities. We first find
\begin{equation}
\begin{split}
\bra{f^0} &H^1 \ket{i^0} \\
&= -\alpha \bra{n}_y y \ket{0}_y \frac{1}{L} \int_{-L/2}^{+L/2} dx \, e^{-i (k_1 - k_0) x} f(x).
\end{split}
\end{equation}
The $\bra{n}_y y \ket{0}_y$ term is zero unless $n = 1$, leading again to the prediction that only $\ket{0}_y \rightarrow \ket{1}_y$ transitions are allowed to first order. Next, we notice that the integral here has the form of a \textit{spatial} Fourier transform: 
\begin{equation}\label{eq:spatial FT}
\begin{split}
\bar{f}(k_x) &= \lim_{L \to \infty} \frac{1}{\sqrt{2\pi}} \int_{-L/2}^{+L/2} dx e^{-i k_x x}  f(x) \\
&= \frac{1}{\sqrt{2\pi}} \int_{-\infty}^{+\infty} dx e^{-i k_x x}  f(x).
\end{split}
\end{equation}
Putting these together, when $n=1$ in $\ket{f^0}$  we find
\begin{equation}
\bra{f^0} H^1 \ket{i^0} = -\frac{\alpha}{L} \sqrt{\frac{\hbar \pi}{\mu \omega_0}} \bar{f}(k_1 - k_0).
\end{equation}

This time-independent expression can be used to calculate the first-order transition coefficient in Eq.~\ref{eq:first-order perturbation}:
\begin{equation*}
c_{k_x = k_{1}, n=1} = \frac{i}{\hbar} \frac{\alpha}{L} \sqrt{\frac{\hbar \pi}{\mu \omega_0}} \bar{f}(k_1 - k_0) \int_{-L/2v}^{+L/2v} dt \, e^{i \omega_{fi} t}.
\end{equation*}
Ultimately, this coefficient should not depend on the box length $L$, since that length was chosen for convenience. But the only way for $L$ dependence to vanish is if $\omega_{fi} = 0$, a condition that forces energy to be conserved as it is exchanged between the beam and the oscillator. 

Presuming that $n=1$ for the oscillator in its final state, setting $\omega_{fi} = 0$ fixes the possible wavenumber $k_1$ for the scattered electron state:
\begin{equation}\label{eq:energy conservation fully quantum}
\begin{split}
\omega_{fi} &= (E_f - E_i)/\hbar = 0\\
\rightarrow 0 &=	 \frac{\hbar}{2 m} \left(k_1^2 - k_0^2 \right) + \omega_0.
\end{split}
\end{equation}
Taking $\omega_{fi} = 0$ also leads to the first-order transition coefficient of 
\begin{equation*}
\begin{split}
c_{k_x = k_{1}, n=1} &= \frac{i}{\hbar} \frac{\alpha}{L} \sqrt{\frac{\hbar \pi}{\mu \omega_0}} \bar{f}(k_1 - k_0) \int_{-L/2v}^{+L/2v} dt \, \cancelto{1}{e^{i \omega_{fi} t}} \\
&= \frac{i}{\hbar} \frac{\alpha}{v} \sqrt{\frac{\hbar \pi}{\mu \omega_0}} \bar{f}(k_1 - k_0),
\end{split}
\end{equation*}
which, squaring, yields the transition probability:
\begin{equation}\label{eq:P1 fully quantum}
P_1 = \frac{\pi \alpha^2}{\hbar \mu \omega_0 v^2} |\bar{f}(k_1 - k_0)|^2.
\end{equation}

This resembles Eq.~\ref{eq:P1, partially quantum} above, and we can show that the two expressions match when $k_1 - k_0$ is small. For our allowed first-order transition, we may write $\omega_{fi} = 0$ as
\begin{equation*}
\begin{split}
0 &= \frac{\hbar}{2 m} (k_1^2 - k_0^2) + \omega_0 \\
\rightarrow 0 &= \frac{\hbar}{2 m} \left[ \cancelto{0}{(k_1 - k_0)^2} + 2 k_0 (k_1 - k_0) \right] + \omega_0,
\end{split}
\end{equation*}
which, if we rearrange and use $k_0 = m v/\hbar$, yields
\begin{equation}
k_1 - k_0 \approx - \omega_0/v,
\end{equation}
which, in turn, allows us to write
\begin{equation}
P_1 \approx \frac{\pi \alpha^2}{\hbar \mu \omega_0 v^2} |\bar{f}(\omega_0/v)|^2
\end{equation}
which, comparing $\bar{f}(k_x)$ in Eq.~\ref{eq:spatial FT} and $\tilde{f}(\omega)$ in Eq.~\ref{eq:temporal FT}, reveals that this $P_1$ indeed matches that of Eq.~\ref{eq:P1, partially quantum}.

\subsection{Numerical Calculation (Setup Only)}

The numerical calculation for the purely quantum model may at first seem like a straightforward generalization of the methods of Sec.~\ref{subsec:partially quantum numerics}. That is, there is nothing to stop one from using the Hamiltonian representation of Eq.~\ref{eq:hamiltonian position basis full quantum}, setting up a large box, and following the time evolution of the Schrodinger equation
\begin{equation}\label{eq:full Schrodinger equation}
i \hbar \frac{\partial}{\partial t} \psi(x,y) = \hat{H}  \psi(x,y)
\end{equation}
using FDTD methods just as above. 

But this calculation is not likely to be terribly informative. Regardless of how sharply peaked the beam particle's spatial wavefunction begins, it will tend to become increasingly broad with time \cite{free_particle_spread}. Understanding the final state, then, will require us to take the spatial Fourier transform of the spread-out wavefunction in $x$ to disentangle probabilities for possible measurements of entangled harmonic oscillator and beam momentum states. 

Of course, we might start out with a partially transformed wavefunction $\bar{\psi}(k_x, y)$ and time evolve that using FDTD. However, the equation of motion for this
\begin{equation}
\begin{split}
i \hbar \frac{\partial}{\partial t} &\bar{\psi}(k_x, y) = \\
 &\left(\frac{\hbar^2 k_x^2}{2 m} - \frac{\hbar^2}{2 \mu} \frac{\partial^2}{\partial y^2} + \frac{1}{2} \mu \omega_0^2 y^2 \right)\bar{\psi}(k_x, y) \\ 
&-\frac{\alpha}{\sqrt{2\pi}} y \bar{f}(k_x)*\bar{\psi}(k_x,y),
\end{split}
\end{equation}
involves a convolution in each time-step
\begin{equation}
\bar{f}(k_x) * \bar{\psi}(k_x,y) =  \int_{-\infty}^{+\infty} dk' \, \bar{f}(k') \, \bar{\psi}(k_x - k', y),
\end{equation}
which is possible, but which (as discussed below) would obscure the $x$-dependence and would make it difficult to distinguish between the ``initial'' and ``final'' states.

\subsection{Approximate Solution}

In the parts of the wavefunction $\psi(x,y)$ where the beam position $x \ll 0$ and $f(x) \approx 0$, we should expect the ``initial'' wavefunction to be essentially that defined by Eq.~\ref{eq:initial state fully quantum}, up to a phase factor. When the beam energy is much greater than the magnitude of the interaction energy, we should expect very little of the beam's incoming wave to be reflected. When the beam reaches the region $x \gg 0$ where again $f(x) \approx 0$, the beam and oscillator once again will evolve without interaction. In this region, the ``final'' wavefunction could be sampled over a large range with $x \gg 0$ and the Fourier transform could be taken in $x$ to disentangle the outcome probabilities. 

To visualize this, we may construct an approximate first-order wavefunction for the scattering states \cite{higherorders}. From our first-order results, we may write \cite{normalization}
\begin{equation}\label{eq:approximate final state wavefunction}
\begin{split}
&\ket{\psi_f(k_x,y,t)} \approx  \\
&\quad\left(\ket{k_0}_x\ket{0}_y + d_1 \sqrt{P_1} \ket{k_1}_x\ket{1}_y \right) e^{-i E_i t/\hbar}
\end{split}
\end{equation}
where $E_i$ is just the system energy established in Eq.~\ref{eq:initial state fully quantum}, $d_1$ is a complex phase factor with $|d_1| = 1$, and the relationship between $k_0$ and $k_1$ is set by Eq.~\ref{eq:energy conservation fully quantum}. 

\begin{figure}
\begin{center}
  \includegraphics[width=.7\columnwidth]{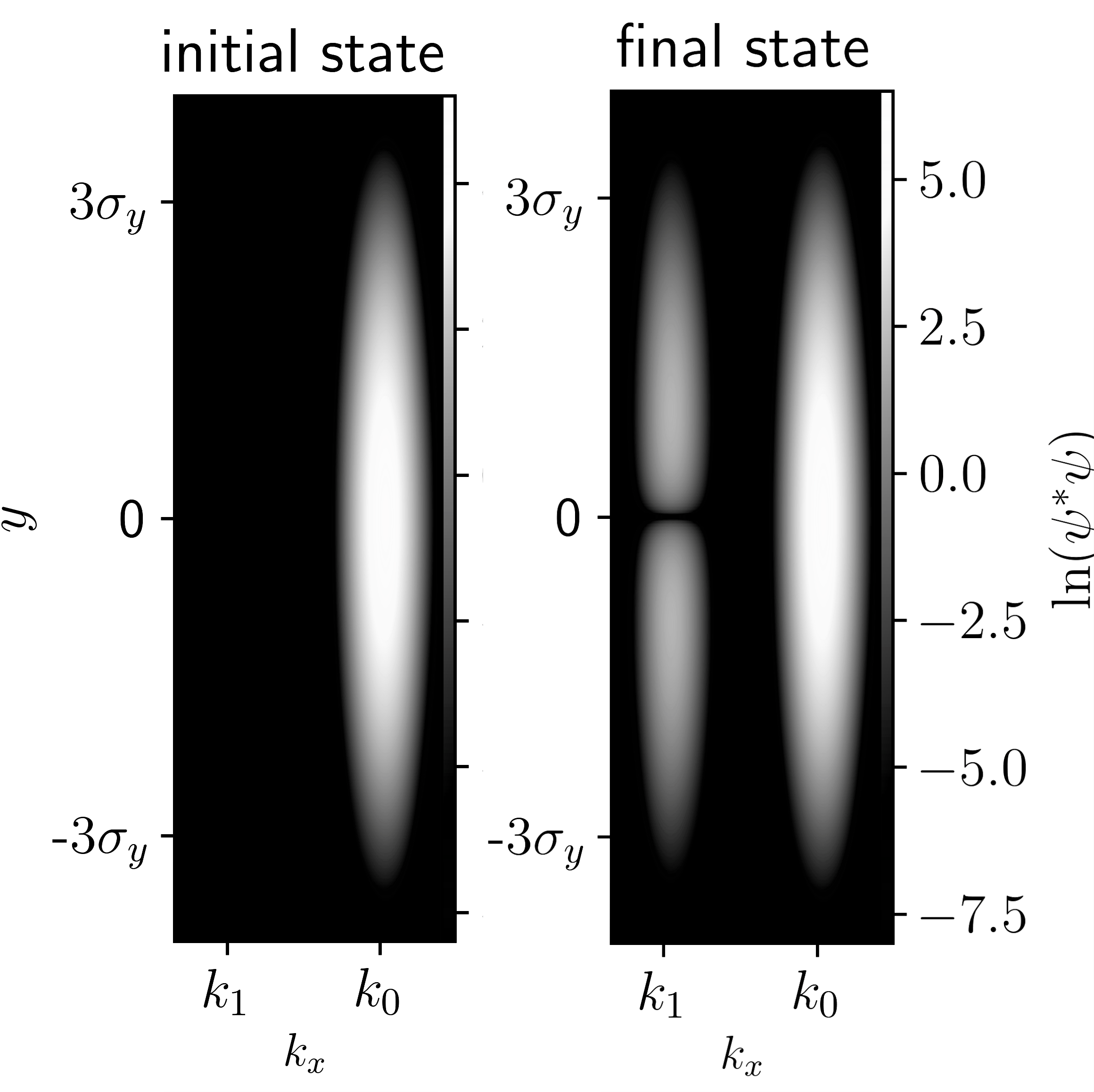}
\end{center}
  \caption{A schematic visualization of the two-particle wavefunction density, before (left) and after (right) scattering.}\label{fig:two-particle wavefunction}
\end{figure}

In this ``final'' region where $x \gg 0$, there will be negligible further interactions between the beam and the oscillator. Hence the probability density of this approximate solution will not evolve further in time:
\begin{equation}
\braket{\psi_f|\psi_f} \approx \delta_{k_x,k_0} |\psi_0(y)|^2 +  P_1 \delta_{k_x,k_1} |\psi_1(y)|^2
\end{equation}
This is plotted in Fig.~\ref{fig:two-particle wavefunction}, using the same parameters as in Fig.~\ref{fig:partially quantum}, albeit with the $k_x$-states broadened for visibility. Visualized in this way, it is easy to imagine these possibilities as distinct ``branches'' of the wavefunction. 

But if this final state density is stationary in time, how would one go down from the higher-level descriptions of $\psi(x,y)$ or $\bar{\psi}(k_x,y)$ to the less informative descriptions of $\psi(y)$ or just the classical $x$ and $y$? This is the essential question of complementarity, to which we now turn.

\section{Complementarity}\label{sec:complementarity}

How might one relate the fully quantum approach to the partially quantum approach? The fully quantum approach of Sec.~\ref{sec:fully quantum approach} uses maximally dispersed plane waves states to describe the beam, while the partially quantum approach of Sec.~\ref{sec:partially quantum approach} treats the beam as a classical particle whose position is sharply defined at all times. Yet the reduced wavefunction $\psi(y)$ should follow
\begin{equation}
\psi(y) = \int_{-\infty}^{+\infty} dx \, \psi(x,y).
\end{equation}
What assumptions might allow this to be true?

Thinking it through more carefully, we can integrate out $x$ on both sides of the two-particle Schrodinger equation (Eq.~\ref{eq:full Schrodinger equation}) to obtain a correct expression for the dynamics of $\psi(y)$, assuming that the wavefunction and its derivatives vanish at infinity. In this case, we find that
\begin{equation}
\begin{split}
i \hbar \frac{\partial}{\partial t} \psi(y) = &-\frac{\hbar^2}{2 \mu}\frac{\partial^2}{\partial y^2}\psi(y) + \frac{1}{2}\mu \omega_o^2 y^2 \psi(y) \\
&- \int_{-\infty}^{+\infty} dx\, \alpha\, y\, f(x)\,  \psi(x,y).
\end{split}
\end{equation}
Comparing this to the Hamiltonian of the partially quantum approach (Eq.~\ref{eq:oscillator hamiltonian partial}), we can see that the only difference will occur in the interaction terms of the two approaches. Setting these terms equal, we find
\begin{equation}
\int_{-\infty}^{+\infty} dx\, f(x)\, \psi(x,y) \approx f(vt) \int_{-\infty}^{+\infty} dx \, \psi(x,y).
\end{equation}

In other words, for the partially and fully quantum approaches to agree, $f(x)$ must both be moved outside the integral, which is reasonable only if $f(x)$ varies much more slowly in $x$ than $\psi(x,y)$, and must also follow $f(x) \approx f(vt)$, which is reasonable only when $x$ is near $\langle x \rangle$, presuming that $\langle x \rangle \approx vt$. Since the region near $\langle x \rangle$ is also where the interaction term will be most significant to the dynamics of the combined wavefunction, such an approximation may be less egregious than it first seems.

On a more limited level, how might we recover the time-dependence of $\psi^*(y)\psi(y)$ when our approximate $\braket{\psi_f|\psi_f}$ appears to be stationary in time? Here we need to integrate out the $\ket{k}_x$ parts of $\ket{\psi(k_x,y,t)}$, including their time dependence. We can do this reduction explicitly for our approximate final-state wavefunction, Eq.~\ref{eq:approximate final state wavefunction}:
\begin{equation}\label{eq:reduced approximate wavefunction}
\begin{split}
\ket{\psi(y,t)} &= \sum_{k_x} \bra{k_x} e^{i E_k t/\hbar} \ket{\psi(k_x,y, t)} \\
&= \ket{0}_y e^{-i \omega_0 t/2} + d_1 \sqrt{P_1} \ket{1}_y e^{-3 i \omega_0 t/2}. 
\end{split}
\end{equation}

An easy way to check that this reproduces the dynamics of $\psi(y)$ as calculated in Sec.~\ref{sec:partially quantum approach} is to obtain the expected value of $y$. For our reduced approximate wavefunction (Eq.~\ref{eq:reduced approximate wavefunction}), we can calculate $\langle y \rangle$ as 
\begin{equation}
\bra{\psi(y,t)}y\ket{\psi(y,t)} = \sqrt{\frac{2 \hbar P_1}{\mu \omega_0}} \mathrm{Re}\left[ d_1 e^{-i \omega_0 t}\right].
\end{equation}
The amplitude of the oscillation for this expected value exactly matches the amplitude of the classical displacement for the oscillator (Eq.~\ref{eq:connection between classical amplitude and quantum probability}), just as it should. 

The single-particle wavefunction $\psi(y)$ can also be linked to the classical oscillator displacement $y$ using Ehrenfest's theorem \cite{Ehrenfest_theorem}. For any quantum operator $\hat{A}$, Ehrenfest's theorem predicts
\begin{equation}\label{eq:generalized Ehrenfest theorem}
\frac{d}{dt} \langle \hat{A} \rangle = \frac{1}{i \hbar} \langle [\hat{A}, \hat{H}] \rangle + \left\langle \frac{\partial \hat{A}}{\partial t} \right\rangle.
\end{equation}

We may apply this machinery to connect the partially quantum approach to the classical approach. Using the Hamiltonian Eq.~\ref{eq:oscillator hamiltonian partial} and the position-momentum commutator Eq.~\ref{eq:commutator}, we can easily calculate
\begin{equation}
[\hat{p}_y, \hat{H}] = +i \hbar \left(-\mu \omega^2 y + \alpha f(vt) \right),
\end{equation}
which can be inserted into Eq.~\ref{eq:generalized Ehrenfest theorem} to yield
\begin{equation}
\frac{d \langle \hat{p}_y \rangle}{dt} = -\mu \omega_0^2 \langle y \rangle + \alpha f(vt).
\end{equation}
This can be compared with the classical equation
\begin{equation}\label{eq:classical oscillator momentum EOM}
\frac{dp_y}{dt} = -\mu \omega_0^2 y + \alpha f(vt),
\end{equation}
demonstrating that $\langle \hat{p}_y \rangle$ and $p_y$ follow the same dynamics. The same exercise can be carried out for $\langle y \rangle$ and $y$, and the first-order equations in time can then be combined to yield second-order equations \textit{a la} Eq.~\ref{eq:classical equations of motion} above.

\section{Entanglement}\label{sec:entanglement}

Schroeder has pointed out that interactions in quantum systems generically introduce entanglement \cite{entanglement_not_spins}. Returning to the approximate final-state wavefunction of Eq.~\ref{eq:approximate final state wavefunction}, we can easily confirm that the state is indeed entangled, since it cannot be written as the product of single-particle wavefunctions. How, then, would our expectations about measurements be altered by the order in which we measure the oscillator and the beam?

In the usual way of discussing quantum mechanics, the wavefunction $\ket{\psi}$ ``collapses'' upon measurement \cite{quantum_paradigms}. The modified wavefunction post-measurement $\ket{\psi'}$ can be written in terms of a projection operator $\hat{\Pi}$ as
\begin{equation}\label{eq:collapse postulate}
\ket{\psi'} = \frac{\hat{\Pi} \ket{\psi}}{\sqrt{\bra{\psi} \hat{\Pi} \ket{\psi}}}.
\end{equation}
The projection operator, here, collapses the wavefunction into a determinate state for whatever relevant quantity has just been measured, and the denominator serves to reinforce normalization on the collapsed wavefunction.

Measuring one particle in a system but not the other requires us to alter the final state wavefunction using a partial projection operator \cite{partial_projection}. For instance, were we first to measure the final momentum of our beam particle with a value $k_1$, the projection operator
\begin{equation}
\hat{\Pi}_{k_1} = \ket{k_1}\bra{k_1}_x \otimes \hat{1}_y, 
\end{equation}
could be applied to our wavefunction approximation (Eq.~\ref{eq:approximate final state wavefunction}) to produce a post-collapse wavefunction of  
\begin{equation}
\ket{\psi'} = \ket{k_1}_x\ket{1}_y,
\end{equation}
where we have omitted the factor of $d_1  e^{-i E_i t/\hbar}$, since this is a phase factor of unit magnitude. Notice, then, that any subsequent predictions for the oscillator would simply match those of its first energy eigenstate.

Likewise, were we first to measure the oscillator's displacement as $y'$, we could update our wavefunction using the projection operator
\begin{equation}
\hat{\Pi}_{y'} = \hat{1}_x \otimes \ket{y'}\bra{y'}_y , 
\end{equation}
which, realizing that $\braket{y'|n}_y = \psi_n(y')$, would lead to the updated wavefunction (again, omitting a phase factor) of
\begin{equation}
\begin{split}
&\ket{\psi'} = \\
&\quad \frac{\psi_0(y') \ket{k_0}_x \ket{y'}_y + d_1 \sqrt{P_1} \psi_1(y') \ket{k_1}_x \ket{y'}_y}{\sqrt{|\psi_0(y')|^2 + |\psi_0(y')|^2}}.
\end{split}
\end{equation}
From this, we find that a measurement of $y'$ far from the origin increases the probability that the beam has reduced its momentum from $k_0$ to $k_1$, since the oscillator's ground-state wavefunction $\psi_0(y)$ (Eq.~\ref{eq:oscillator normalized ground state}) has thinner tails than the excited-state wavefunction $\psi_1(y)$ (Eq.~\ref{eq:oscillator normalized first excited state}). 

\begin{figure}
\begin{center}
  \includegraphics[width=\columnwidth]{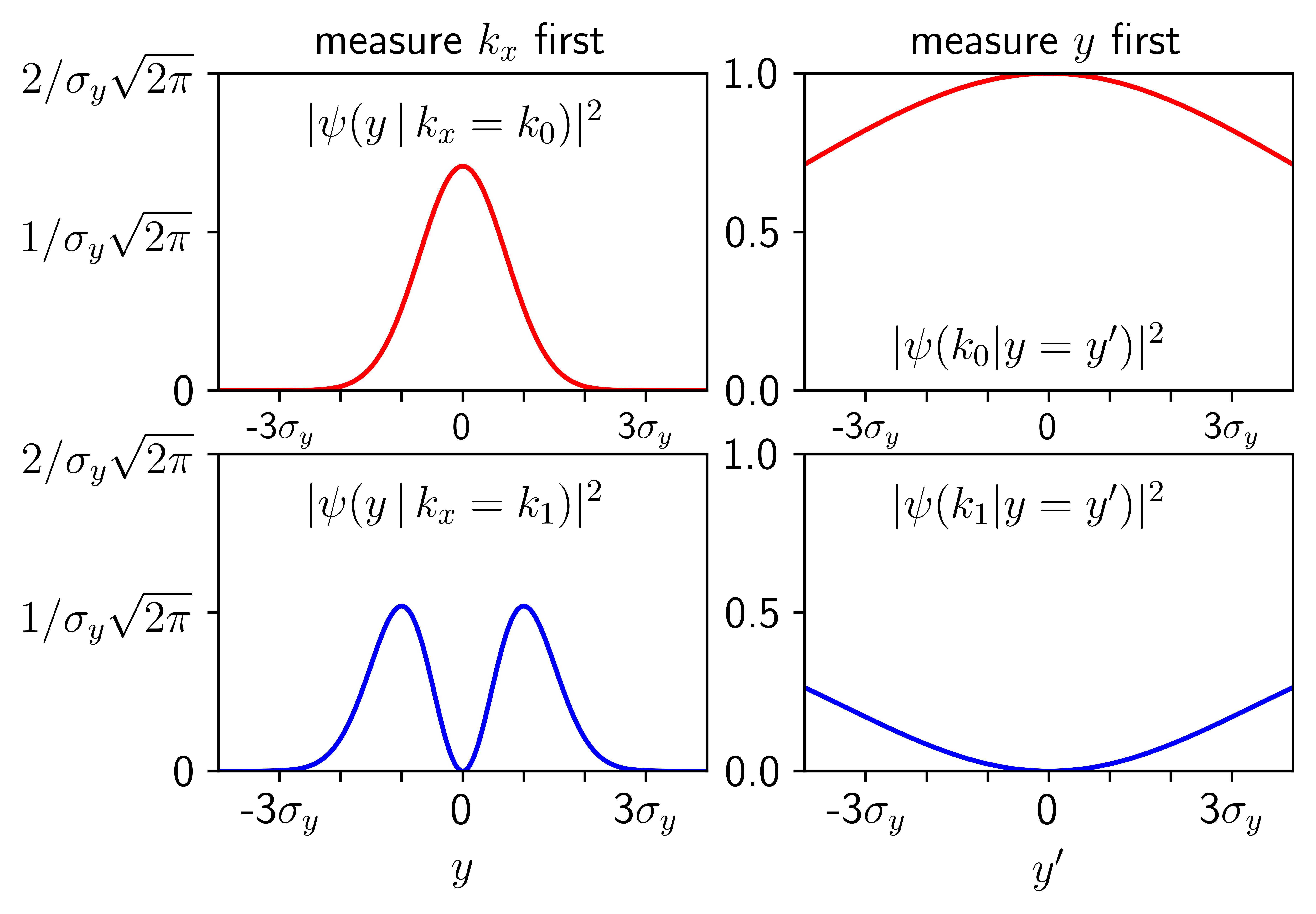}
\end{center}
  \caption{\textit{Left}: If the beam momentum is measured first as $\hbar k_0$ (top left), the oscillator displacement predictions can be found by integrating over the ground-state wavefunction $|\psi_0(y)|^2$. If the beam momentum is measured first as $\hbar k_1$ (bottom left), the oscillator displacement predictions can be found by integrating over the excited-state wavefunction $|\psi_1(y)|^2$. \textit{Right}: Given a measurement of oscillator displacement $y'$, smaller magnitudes of $y'$ increase the probability of measuring the beam in its original momentum state $\hbar k_0$ (top right), and larger magnitudes of $y'$ increase the probability of measuring the beam in its reduced momentum state $\hbar k_1$ (bottom right).}\label{fig:entanglement illustration}
\end{figure}

These results are summarized in Fig.~\ref{fig:entanglement illustration}, which uses all the same model parameters as in Figs.~\ref{fig:partially quantum} and~\ref{fig:two-particle wavefunction}. If the beam momentum is measured first, this changes our predictions for the oscillator displacement probabilities. If the oscillator displacement is measured first, this changes our predictions for the beam momentum probabilities. 

While such claims are undoubtedly of academic interest, we should concede how minimally they bear on most real experiments. To make an analogy with real experiments, the ``oscillator'' is just the sample being studied, and the ``beam'' is just the probe being used. Given the results of Fig.~\ref{fig:entanglement illustration}, why is the order of quantum measurements not a critical part of all experimental protocols? 

It is because any ``measurement'' typically involves further interactions, and thus further branching and entanglement. The entanglement of the sample with the probe allows experimental measurements of the probe to tell us something about the sample \cite{inelastic entanglement}, but it typically is difficult to make further measurements on the sample that preserve quantum coherence. After all, most quantum systems whose positions can be pinned down (e.g., an atomic defect in a crystal) interact with their substrates and the beam alike, and the intrusions of environmental decoherence conspire to obviate the need for such worries.\cite{decoherence}

\section{Conclusion}\label{sec:conclusion}

What, then, has been learned? Three complementary approaches to a simple model Hamiltonian yield results that conceptually differ, but quantitatively match. In the classical approach, the beam particle transfers a predictable quantity of energy to the oscillator as it passes. When the classical oscillator is replaced by a quantum oscillator, the quantum oscillator's displacement expectation value oscillates with an amplitude matching that of the classical oscillator, and the quantum amplitude of the first excited state has a magnitude that is directly proportional to the classical amplitude. When both the beam particle and the harmonic oscillator are treated as quantum objects, however, the interaction between the two objects induces entanglement. The conditions allowing these three approaches to match were explored, and the consequences of partial wavefunction ``collapse'' on measurements of entangled systems were demonstrated. 

\,
\section*{ACKNOWLEDGMENTS}

The anonymous reviewers and the journal editor offered helpful suggestions that improved this manuscript.

\section*{AUTHOR DECLARATIONS}

\subsection*{Conflict of Interest}

The author has no conflicts of interest to disclose.

\subsection*{Code Availability}

The Python scripts used to generate figures are available in the Supplemental Materials.

\appendix   

\section{Classical Work Calculation}

To calculate the classical work done on the beam as it passes the harmonic oscillator, one can rewrite the work integral as an integral over time
\begin{equation*}
W_\mathrm{beam} = \int_{-\infty}^{+\infty} dx \cdot F_x = \int_{-\infty}^{+\infty} v dt \left[ \alpha y(t) \frac{1}{v}\frac{df}{dt} \right],
\end{equation*}
insert $y(t)$ (from $y(\omega)$ in Eq.~\ref{eq:classical FT of y}) as an inverse Fourier transform
\begin{equation*}
W_\mathrm{beam} = \alpha \int_{-\infty}^{+\infty} dt \left[ \left( \frac{1}{\sqrt{2\pi}} \int_{-\infty}^{+\infty} d\omega e^{-i\omega t} \frac{-\alpha \tilde{f}(\omega)}{\mu (\omega^2 - \omega_0^2)} \right) \frac{df}{dt} \right],
\end{equation*}
\newline
and reverse the order of the time and frequency integrals:
\begin{equation*}
W_\mathrm{beam} = -\frac{\alpha^2}{\mu} \int_{-\infty}^{+\infty} d\omega \frac{\tilde{f}(\omega)}{(\omega^2 - \omega_0^2)} \frac{1}{\sqrt{2\pi}}\int_{-\infty}^{+\infty} dt \, e^{-i\omega t} \frac{df}{dt}.
\end{equation*}
The time integral can be made into the complex conjugate of a Fourier transform after an integration by parts, and the frequency integral may be performed by slightly displacing the poles off the real axis by $i \epsilon$:
\begin{equation*}
W_\mathrm{beam} = -\frac{\alpha^2}{\mu}\int_{-\infty}^{+\infty} d\omega \frac{\tilde{f}(\omega)(-i \omega) \tilde{f}^*(\omega)}{(\omega - \omega_0 +i\epsilon) (\omega + \omega_0 - i\epsilon)}.
\end{equation*}
Using the residue theorem, we find
\begin{equation}
W_\mathrm{beam} =-\frac{\pi \alpha^2}{\mu} \left| \tilde{f}(\omega_0) \right|^2.
\end{equation}
This is the expression reported above as Eq.~\ref{eq:work on beam}.


\begin{thebibliography}{99}
\bibitem{higherorders} The ``partially quantum'' approach of Sec.~\ref{sec:fully quantum approach} is simple enough that it can be solved exactly, as is done in Lucas and Šunjić \cite{original_reference}, yielding probabilities for oscillator energy eigenstates $\ket{n}$ of $e^{-P_1} P_1^n/n!$ (i.e., the probabilities form a Poisson distribution). Higher-order contributions to a scattering wavefunction could be constructed using this, but little would be gained beyond algebraic clutter.
\bibitem{normalization} The approximate wavefunction of Eq.~\ref{eq:approximate final state wavefunction} is nearly normalized since $P_1$ is small, but readers who would like strict normalization may introduce a $\sqrt{1-P_1}$ before the first term. Again, little would be gained.
\end{thebibliography}
\end{document}